\documentstyle[12pt]{article}
\begin{document}
\begin{center}
{\large\bf Exact Renormalization Group and Running Newtonian Coupling in
Higher Derivative Gravity\\}
\vspace{1cm}
A.A. Bytsenko
\footnote{e-mail: abyts@spin.hop.stu.neva.ru}\\
{\it State Technical University, St. Petersburg, 195252, Russia}\\

L.N. Granda
\footnote{e-mail: granda@quantum.univalle.edu.co}\\
{\it Departamento de Fisica, Universidad del Valle\\
A.A. 25360, Cali, Colombia}\\

S.D. Odintsov
\footnote{e-mail: odintsov@quantum.univalle.edu.co}\\
{\it Tomsk Pedagogical University, 634041 Tomsk, Russia\\
and\\
Departamento de Fisica, Universidad del Valle\\
A.A. 25360, Cali, Colombia}\\
\vspace{1.5cm}
{\bf Abstract}

We discuss exact renormalization group (RG) in $R^2$-gravity using effective 
average action formalism. The truncated evolution equation for such a theory 
on De Sitter background leads to the system of non perturbative RG equations 
for cosmological and gravitational coupling constants.\\
Approximate solution of these RG equations  shows the appearence of 
antiscreening and screening behaviour of Newtonian coupling what depends on 
higher derivative coupling constants.\\
\end{center}
In the absence of consistent quantum gravity it could be that consideration of
effective models  for quantum gravity (QG) is the only possibility to take 
into account gravitational phenomena in high energy physics. One may start from
particular model of QG (see \cite{bos} for a review) and to formulate effective
model which describes theory in some region. In such a way, effective theory 
for conformal factor to describe QG in far infrared (at large distances) has 
been formulated \cite{am}. Such theory which is based on higher derivative 
scalar gives the way to estimate the behaviour of Newtonian coupling 
\cite{ao}.

One may consider Einstein gravity as effective theory and estimate quantum
corrections to Newtonian coupling \cite{hl} using effective field theory
technique. Moreover, as non-renormalizability is not a problem in such
approach one can apply exact RG \cite{wk}, say in a form of effective average
action, in order to formulate the non perturbative RG equations for coupling 
constants in Eistein gravity \cite{re,fo}.
In the same way it is very interesting to consider $R^2$-gravity as effective 
model. Such model attracts a lot of attention (see [1] for a review and list 
of references), being multiplicatively renormalizable (but eventually 
non-unitary in perturbative approach). Note that perturbative RG equations for
higher derivative gravity have been first considered in ref.\cite{jt} 
(see [1] for an introduction). Kind of effective $R^2$-gravity leads to more 
or less successful unflationary Universe \cite{st}.

In the present letter we formulate the evolution equation and non-perturbative
RG equations for coupling constants in $R^2$-gravity [1].
The action to start with is given by the following (in Euclidean notations)
\par$$
S=\int d^4x \sqrt{g} \left\{\epsilon R^*R^*+\frac{1}{2f^2}C_{\mu\nu\alpha\beta}
C^{\mu\nu\alpha\beta}-\frac{1}{6\nu^2}R^2-2\kappa^2 R+4\kappa^2 \Lambda\right\}
\mbox{,}
\eqno{(1)}$$
where $R^*R^*=\frac{1}{4}\epsilon^{\mu\nu\alpha\beta}\epsilon_{\lambda\rho
\gamma\delta}R^{\lambda\rho}_{\mu\nu}R^{\gamma\delta}_{\alpha\beta}$, $C_
{\mu\nu\alpha\beta}$ is the Weyl tensor, $\kappa^{-2}=32\pi\bar{G}$ is the 
Newton constant, $\epsilon, f^2,
\nu^2$ are gravitational coupling constants. It is well known that the theory 
with action (1) is multiplicatively renormalizable and asymptotically free. 
Note that perturbative running of Newtonian coupling constant in a theory (1) 
with matter has been disscussed in ref. \cite{eld}.

Following the approach of ref. \cite{re} we will write the evolution equation for 
the effective average action $\Gamma_k[g,\bar{g}]$ defined at non zero momentum
ultraviolet scale $k$ below some cut-off $\Lambda_{cut-off}$. The truncated 
form of such evolution equation has the following form:
\par$$
\partial_t\Gamma_k[g,\bar{g}]=\frac{1}{2}Tr\left[\left(\Gamma_k^{(2)}[g,
\bar{g}]+R_k^{grav}[\bar{g}]\right)^{-1}\partial_tR_k^{grav}[\bar{g}]\right]-
$$
$$
\sum_i c_iTr\left[\left(-M_i[g,\bar{g}]+R_{ki}^{gh}[\bar{g}]\right)\partial_tR_
{ki}^{gh}[\bar{g}]\right]\mbox{,}
\eqno{(2)}$$
here $t=ln k$, $R_k$ are cutoffs in gravitational and ghosts sectors, $c_i$ 
are the weights for ghosts (we have Fadeev-Popov ghost with $c_{FP}=1$ and  
so called third ghost with weight $c_{TG}=1/2$),
$g_{\mu\nu}=\bar{g}_{\mu\nu}+h_{\mu\nu}$ where $h_{\mu\nu}$ is the quantum  
gravitational field, $\Gamma_k^{(2)}$ is the Hessian of $\Gamma_k[g,\bar{g}]$ 
with respect to $g_{\mu\nu}$ at fixed $\bar{g}_{\mu\nu}$, $M_i$ are ghosts 
operators. Note that the RHS of eq. (2) is very similar to the one-loop 
effective action.

At the next steep  we have to specify the truncated evolution equation for the
theory (1). Starting from UV scale $\Lambda_{cut-off}$, evolving the theory 
down to smaller scales $k<<\Lambda_{cut-off}$ we may use the truncation of the
form
\par$$
\kappa^2\rightarrow Z_{Nk}^{-1}\kappa^2\;\;\mbox{,}\frac{1}{f^2}\rightarrow Z_
{Nk}\frac{1}{f^2}\;\;\mbox{,} \frac{1}{\nu^2}\rightarrow Z_{Nk}\frac{1}{\nu^2}
\;\;\mbox{,} \Lambda\rightarrow \lambda_k\mbox{,}
\eqno{(3)}$$
where $k$-dependence is denoted by index $k$.
We will be limited here only to lower derivatives terms in the reduction of 
$\Gamma_k$, i.e. higher derivatives coupling constants may be considered as 
free parameters.

Choosing $\bar{g}_{\mu\nu}=g_{\mu\nu}$ (then ghost term disappears) and 
projecting the evolution equation on the space with low derivatives terms one 
gets the left-hand side of the evolution eq. (2) as following:
\par$$
\partial_t\Gamma_k[g,g]=2\kappa^2\int d^4x \sqrt{g}\left[-R(g)\partial_tZ_{Nk}
+2\partial_t\left(Z_{Nk}\lambda_k\right)\right]\mbox{.}
\eqno{(4)}$$
The initial conditions for $Z_{Nk}$, $\lambda_k$ are choosen as in ref. \cite{re}.

The right-hand side of evolution equations may be found after very tedious 
calculations (choosing De Sitter background $R_{\mu\nu}=1/4 g_{\mu\nu}R$, 
calculating the path integral, making expansion on $R$). We drop the details of 
these calculations.
The final system of non-perturbative RG equations for Newtonian and 
cosmological constants is obtained as following:
\par$$
\partial_t g_k=\left[2+\eta_N(k)\right]g_k\mbox{,}
\eqno{(5)}$$
where $g_k$ is the dimensionless renormalized Newtonian constant,
\par$$
g_k=k^2G_k=k^2Z_{Nk}^{-1}\bar{G}\mbox{.}
\eqno{}$$
The anomalous dimension $\eta_N(k)$ is given by
\par$$
\eta_N(k)=g_kB_1(\alpha_{2k},\beta_{2k},\gamma_{2k},\delta_{2k})+\eta_N(k)g_
kB_2(\alpha_{2k},\beta_{2k},\gamma_{2k},\delta_{2k})\mbox{,}
\eqno{(6)}$$
where
\par$$
B_1(\alpha_{2k},\beta_{2k},\gamma_{2k},\delta_{2k})=\frac{1}{12\pi}\left\{10
\Phi_1^1(\alpha_{2k})+10\Phi_1^1(\beta_{2k})\right.
$$
$$
\left.-10\Phi_1^1(0)+2\Phi_1^1(\gamma_{2k})+2\Phi_1^1(\delta_{2k})-\left(60
\alpha_1+5\right)\Phi_2^2(\alpha_{2k})-\left(60\beta_1+5\right)\Phi_2^2(\beta_
{2k})\right.
$$
$$
\left. +(\frac{24}{K-3}-6)\Phi_2^2(0)-12\gamma_1\Phi_2^2(\gamma_{2k})-12
\delta_1\Phi_2^2(\delta_{2k})\right\}\mbox{,}
\eqno{}$$
\par$$
B_2(\alpha_{2k},\beta_{2k},\gamma_{2k},\delta_{2k})=-\frac{1}{12\pi}\left\{5
\tilde{\Phi}_1^1(\alpha_{2k})+5\tilde{\Phi}_1^1(\beta_{2k})+7\tilde{\Phi}_1^1
(0)\right.
$$
$$
\left.+\tilde{\Phi}_1^1(\gamma_{2k})+\tilde{\Phi}_1^1(\delta_{2k})-30(\alpha_1
+\frac{1}{12})\tilde{\Phi}_2^2(\alpha_{2k})-30(\beta_1+\frac{1}{12})\tilde{
\Phi}_2^2(\beta_{2k})\right.
$$
$$
\left. -3\tilde{\Phi}_2^2(0)-6\gamma_1\tilde{\Phi}_2^2(\gamma_{2k})-6\delta_1
\tilde{\Phi}_2^2(\delta_{2k})\right\}\mbox{.}
\eqno{(7)}$$
Here
\par$$
\alpha_1\mbox{,}\beta_1=\frac{1}{12}+\frac{f^2+\nu^2}{6\nu^2}\pm\frac{1}{2}
\left(\frac{f^2+\nu^2}{3\nu^2}-\frac{K+6}{6K}\right)\left[1+\frac{4\lambda_k}
{\kappa^2f^2K}\right]^{-1/2}\mbox{,}
\eqno{}$$
\par$$
\gamma_1\mbox{,}\delta_1=\frac{1}{2(K-3)}\pm \frac{1}{2(K-3)}\left(1-\frac{8
\lambda_k}{\kappa^2\nu^2K}\right)^{-1/2}\mbox{,}
\eqno{(8)}$$
\par$$
\alpha_{2k}\mbox{,}\beta_{2k}=\frac{\kappa^2 f^2}{k^2}\left\{1\pm\left[1+\frac
{4\lambda_k}{\kappa^2 f^2K}\right]^{1/2}\right\}\mbox{,}
\eqno{}$$
\par$$
\gamma_{2k}\mbox{,}\delta_{2k}=\frac{\kappa^2\nu^2}{k^2}\left\{1\pm \left[1-
\frac{8\lambda_k}{\kappa^2\nu^2(K-3)}\right]^{1/2}\right\}\mbox{.}
\eqno{(9)}$$
Note that $K=\frac{3f^2}{f^2+2\nu^2}$ what corresponds to the choice of 
so-called gauge-fixing independent effective action (for a review see 
\cite{bos,sd}). By this choice, we solve the gauge-dependence problem
(for a related discussion in case of Einstein gravity, see \cite{fo}).
The functions $\Phi_n^p(w)$ and $\tilde{\Phi}_n^p$ are given by the integrals
\par$$
\Phi_n^p(w)=\frac{1}{\Gamma(n)}\int_0^{\infty}dz z^{n-1}\frac{R^{(0)}(z)-zR^
{(0)'}(z)}{[z+R^{(0)}(z)+w]^p}\mbox{,}
$$
$$
\tilde{\Phi}_n^p(w)=\frac{1}{\Gamma(n)}\int_0^{\infty}dz z^{n-1}\frac{R^{(0)}
(z)}{[z+R^{(0)}(z)+w]^p}\mbox{.}
\eqno{(10)}$$

Solving (6)
\par$$
\eta_N(k)=\frac{g_kB_1(\bar{\lambda}_k,\kappa_k)}{1-g_kB_2(\bar{\lambda}_k,
\kappa_k)}\mbox{,}
\eqno{(11)}$$
where $\kappa_k^2=\kappa^2/k^2$, $\bar{\lambda}_k=\lambda_k/k^2$,
we see that the anomalous dimension $\eta_N$ is a non pertubative quantity. 
The evolution equation for the cosmological constant is obtained as following
\par$$
\partial_t(\bar{\lambda}_k)=-\left[2-\eta_N(k)\right]\bar{\lambda}_k+
\frac{g_k}{4\pi}\left\{10\Phi_2^1(\alpha_{2k})+10\Phi_2^1(\beta_{2k})-10
\Phi_2^1(0)\right.
$$
$$
\left.
+2\Phi_2^1(\gamma_{2k})+2\Phi_2^1(\delta_{2k})-\eta_N(k)\left[5\tilde{\Phi}_2^
1(\alpha_{2k})+5\tilde{\Phi}_2^1(\beta_{2k})\right.\right.
$$
$$
\left.\left.+7\tilde{\Phi}_2^1(0)+\tilde{\Phi}_2^1(\gamma_{2k})+\tilde{\Phi}_
2^1(\delta_{2k})\right]\right\}\mbox{.}
\eqno{(12)}$$
Eqs. (5) and (12) with (11) determine the value of the running Newtonian 
constant and cosmological constant at the scale $k<<\Lambda_{cut-off}$. Above 
evolution eqs. include non-perturbative effects which go beyond a simple 
one-loop calculation.

Next we estimate the qualitative behaviour of the running Newtonian constant 
as above system of RG equations is too complicated and cannot be solved 
analytically. To this end we assume that the cosmological constant is much 
smaller than the IR cut-off scale, $\lambda_k<<k^2$, so we can put 
$\lambda_k=0$ that simplify Eqs. (8) and (9). After that, we make an expansion
in powers of $(\bar{G}k^2)^{-1}$ keeping only the first term (i.e. we evaluate 
the functions $\Phi_n^p(0)$ and $\tilde{\Phi}_n^p(0)$) and
finally obtain (with $g_k\sim k^2\bar{G}$)
\par$$
G_k=G_o\left[1-w\bar{G}k^2+...\right]\mbox{,}
\eqno{(13)}$$
where
\par$$
w=-\frac{1}{2}B_1(0,0)=\frac{1}{24\pi}\left[\left(50+22\frac{f^2}{\nu^2}
\right)-\frac{7\pi^2}{3}\right]\mbox{.}
\eqno{}$$
In case of Einstein gravity, similar solution has been obtained in refs. \cite{re,fo}.
In getting (13) we use the same cut-off function as in \cite{re}.

We see that sign of $w$ depends on higher derivatives coupling constants:
\par$$
w>0 \;\;\mbox{,if}\;\;\ 50-\frac{7\pi^2}{3}+\frac{22f^2}{\nu^2}>0\mbox{.}
\eqno{(14)}$$
The coupling constant $\nu^2$ maybe choosen to be negative (see [1]). So, for 
example for $f^2=1$, $\nu^2=\pm 1$ we get $w>0$ and Newtonian coupling 
decreases as $k^2$ increases; i.e. we find that gravitational coupling is 
antiscreening. On the contrary, for $f^2=1$, $\nu^2=-1/2$ we get $w<0$ and 
screening behaviour for Newtonian coupling. It means that in such phase 
gravitational charge (mass) is screened by quantum fluctuations, or, in other 
words Newtonian coupling is smaller at smaller distances. The sign of quantum 
correction to Newtonian potential will be different also.

Note that above quantum correction to Newtonian coupling constant has been 
calculated in ref.[10] using one-loop approach and perturbative RG equations. 
It is clear that result of such calculation is different from the one presented
above as we use nonperturbative RG method. Moreover, as it has been noted at 
the beginning the theory under discussion is multiplicatively renormalizable 
in perturbative approach, but most likely it is not unitary in such approach. 
Hence, the perturbative results may not be trusted in many situations. On the 
contrary, within nonperturbative approach the theory is considered as an 
effective theory, so the problems with non-unitarity are not important. 
The possibility to get some nonperturbative results in models of QG 
in four dimensions looks very attractive and may help in the construction of 
new QG models.

Thus, we found that Newtonian coupling may show screening or antiscreening 
behaviour in $R^2$-gravity what depends from higher derivative couplings. That
shows explicitly that $R^2$ quantum gravity may lead to different physical 
consequences than Einstein gravity even at low energies.

Acknowledgments.This work has been supported by COLCIENCIES(Colombia),
GRASENAS 95-0-6.4-1 and by RFBR,Project No.96-02-16017.


\begin{thebibliography}{99}

\bibitem{bos} I.L. Buchbinder, S.D. Odintsov and I.L. Shapiro, Effective 
Action in Quantum Gravty, IOP Publishing, Bristol, 1992

\bibitem{am} I. Antoniadis and E. Mottola, Phys. Rev. {\bf D 45} (1992), 2013;
S.D. Odintsov, Z. Phys. {\bf C 54} (1992), 531

\bibitem{ao} I. Antoniadis and S.D. Odintsov, Phys. Lett. {\bf B 343} (1995), 
76; E. Elizalde, S.D. Odintsov and I.L. Shapiro, Class. and Quantum Grav. 
{\bf 11} (1994), 1607

\bibitem{hl} H.W. Hamber and S. Liu, Phys. Lett. {\bf B 357} (1995), 51

\bibitem{wk} K.G. Wilson and J. Kogut, Phys. Rep. {\bf 12} (1974), 75;
J. Polchinski, Nucl. Phys. {\bf B 231} (1984), 269

\bibitem{re} M. Reuter, hep-th/9605030 (1996)

\bibitem{fo} S. Falkenberg and S.D. Odintsov, hep-th/9612019 (1996)

\bibitem{jt} J. Julve and M. Tonin, Nuovo Cim. {\bf B 46} (1978), 137

\bibitem{st} A. Starobinsky, Phys. Lett. {\bf B 91} (1980), 99

\bibitem{eld} E. Elizalde, C. Lousto, S.D. Odintsov and A. Romeo, Phys. Rev. 
{\bf D 52} (1995), 2202

\bibitem{sd} S.D. Odintsov, Fortschr. Phys. {\bf 38} (1990), 371

\end{thebibliography}
\end{document}